    \documentclass[12pt]{article} 
    \usepackage{epsfig}  \usepackage{graphics}
     \newlength{\dinwidth}                       
     \newlength{\dinmargin}                      
\def\PLB{{\em Phys. Lett.}  B}
\def\PRL{{\em Phys. Rev. Lett.}}
\def\PRD{{\em Phys. Rev.} D}

\def\EPJC{{\em Eur. Phys. J.} C}
\def\lsim{\mathrel{\rlap{\lower4pt\hbox{\hskip1pt$\sim$}}
    \raise1pt\hbox{$<$}}}                
\def\gsim{\mathrel{\rlap{\lower4pt\hbox{\hskip1pt$\sim$}}
    \raise1pt\hbox{$>$}}}                
\newcommand{\gp}{\mbox{$g_{1}^{\rm p}$}}
\newcommand{\gn}{\mbox{$g_{1}^{\rm n}$}}

\begin{document}
\vspace*{10mm}
\begin{center}  \begin{Large} \begin{bf}
The Physics Case for Polarized Protons at HERA \\ 
\end{bf}  \end{Large}
  \vspace*{5mm}
  \begin{large}
A. Deshpande$^{a}$  
\end{large}

$^a$ Yale University, Physics.\ Dept., New Haven, CT 06520, U.S.A.

\end{center}
\begin{quotation}
\noindent
{\bf Abstract:}
Several important and unique measurements,
within the standard model and of possible physics beyond it, could be made
with {\em polarized HERA} in which both the proton and the electron
beams are polarized. With a $\sim$820 GeV proton beam and a $\sim$27.6 GeV 
electron beam, the polarized HERA will enable $\vec{e}-\vec{p}$ 
collisions with $\sqrt{s} \sim 300$ GeV and access spin variables in 
the kinematic range, $10^{-5} \le x_{\rm Bj} \le 0.6$ 
and $0 \le Q^{2} \le 10^{5}$, using the
H1 and ZEUS detectors at DESY.
This will be an increase of two orders of magnitude 
in both $x$ and $Q^{2}$ range compared to the presently
explored range from fixed target experiments at CERN, SLAC and DESY. 
No other approved or planned spin experiment or accelerator facility 
will access the low $x$ and high $Q^{2}$ regions possible with HERA.
Measurements performed with the polarized HERA collider will include 
the polarized structure function $g_{1}(x,Q^{2})$
at very low $x$, 
the polarized gluon distribution $\Delta G(x,Q^{2})$ 
   from pQCD analysis of $g_{1}$, 
   from the production of di-jet events and high-p$_{T}$ hadrons 
       in photon gluon fusion process and in photoproduction, 
   weak structure functions,
   valence quark distribution functions from semi-inclusive asymmetries, 
   parton distributions inside polarized photon, 
   and 
   information on helicity structure of possible new physics beyond 
       the standard model.
With such a rich and broad physics program possible for HERA,
not polarizing the proton beam would be a great opportunity lost.
\end{quotation}

\section{Polarized DIS in a New {\boldmath $x-Q^{2}$} Region}
\label{sec:intro}
Measurements of nucleon structure functions by lepton-nucleon deep
inelastic scattering (DIS) were of fundamental importance in studying
nucleon structure and have provided crucial information regarding the
foundation, and later, the development of perturbative QCD (pQCD). Historically
in this field important new information has been obtained when experimental
measurements were extended to new kinematic regions\cite{rgr90}. For example,
measurement of elastic scattering 
in the $Q^{2}$ range $\sim 1$ GeV$^{2}$ established the finite size of the
proton. 
The extension of measurements of inelastic
inclusive electron scattering to deep inelastic region $Q^{2} > 1$ GeV$^{2}$
revealed the existence of partonic substructure of the proton, which was later 
studied using muon-nucleon scattering. The significant increase
in the $Q^{2}$ accessible range, especially at low $x$, made possible by 
the $e-p$ collisions 
at HERA revealed the surprising rise in the $F_{2}$ at low $x$ 
whose measurement and analysis in the past 5 years has contributed greatly 
to our understanding of pQCD.

\begin{figure}[here]
\hfil
\epsfxsize=9.0cm
\epsffile{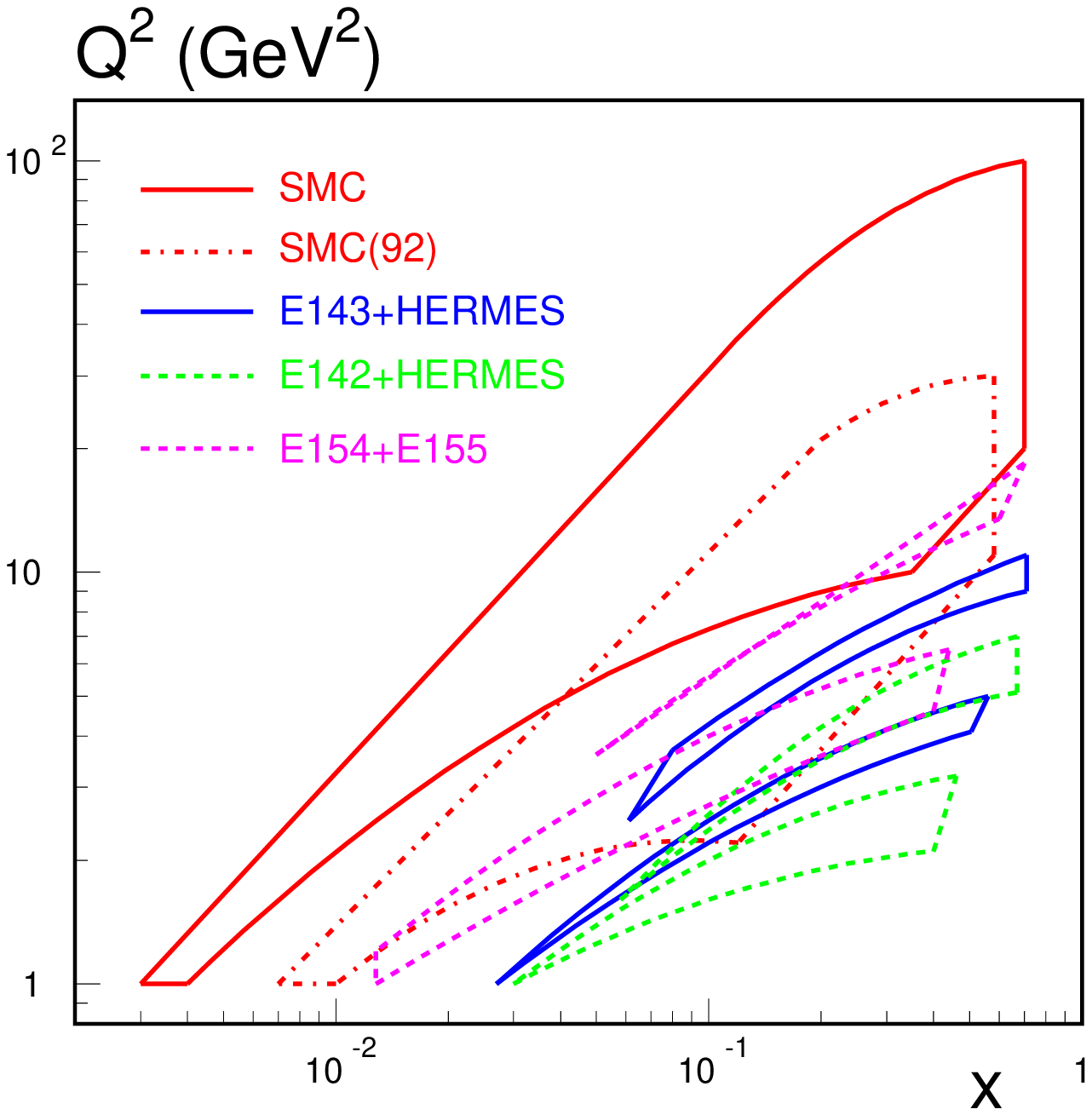}
\epsfxsize=9.0cm
\epsffile{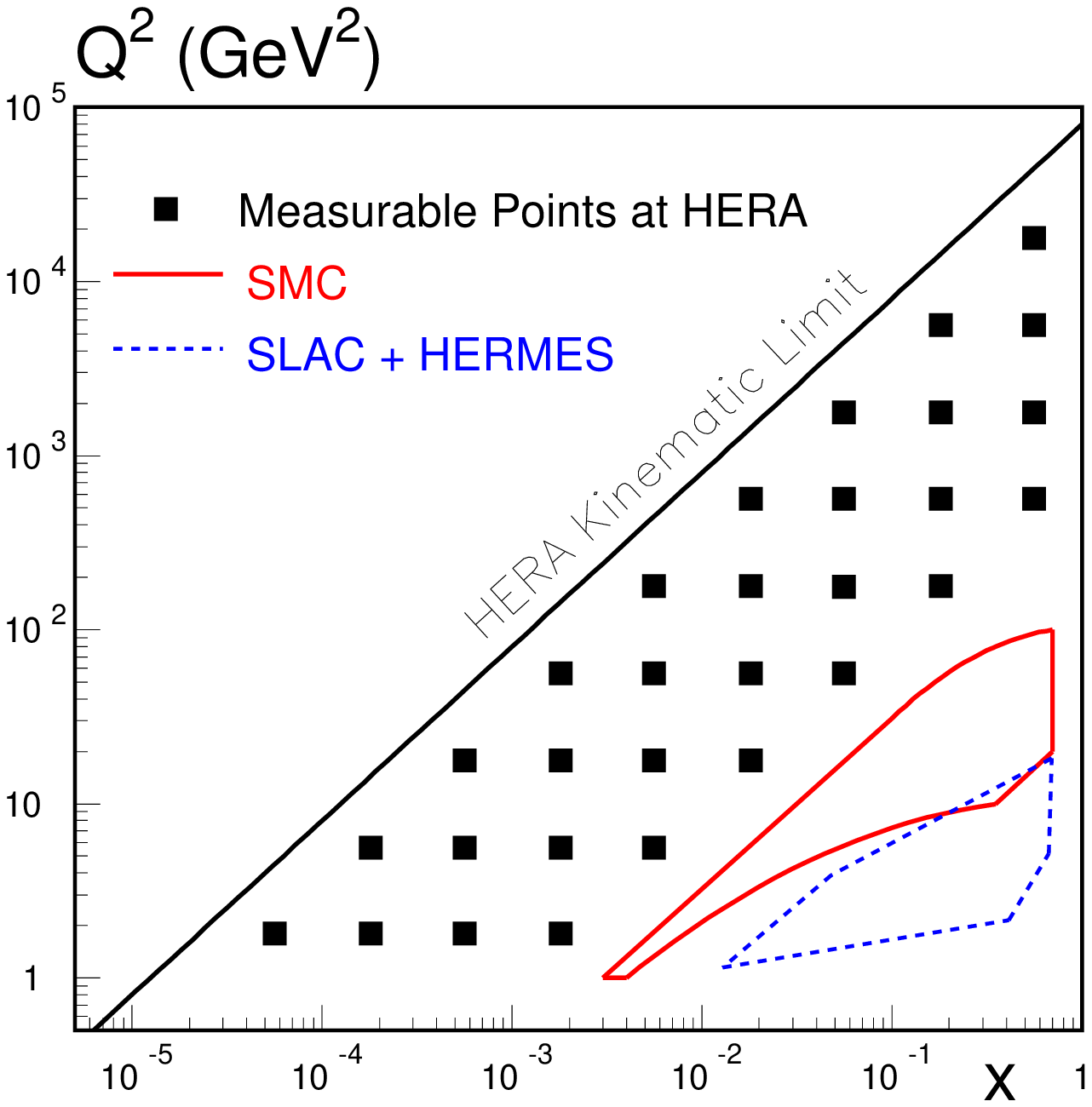}
\hfil
\caption{\em Top: The $x-Q^{2}$ range explored by the CERN, SLAC and DESY fixed target
experiments today. Bottom: the $x-Q^{2}$ range accessible by polarized HERA with 
H1 or ZEUS collider detectors compared with the fixed target experiments.}
\label{fig:xq2}
\end{figure}

A similar historical path has been traced by events in
{\em polarized} DIS\cite{gab99}. First measurements of polarized DIS by the 
Yale-SLAC collaboration
extended the $x$ range down to $x\sim 0.1$ and were consistent with the 
accepted quark-parton model of the proton spin at the time.
Extension of the $x-Q^{2}$ of spin DIS
measurements at CERN by the EMC collaboration (down to $x=0.01$) revealed
that the Ellis-Jaffe Sum rule was violated and the quarks contributed 
hardly anything towards the proton's spin. This surprising result
led to a large number of experimental measurements in the past 10 years
performed by the SMC Collaboration at CERN, a series of measurements by 
E142, E143, E154 and E155 Collaborations at SLAC, and recently by HERMES
Collaboration at DESY. These were fixed target experiments 
with electron beam energy between $\sim 26-49$ GeV at SLAC and DESY which
explored the $\sqrt{s}$ range of $\sim 7-10$ GeV, while SMC used $\sim 200$ GeV 
polarized muon beam with $\sqrt{s} \sim 20$ GeV.
The $x-Q^{2}$ range covered by these experiments {\em together} is: $0.003 \le x \le 0.7$
and $0.2 \le Q^{2} \le 100$ GeV$^{2}$ (figure \ref{fig:xq2}, Top). All 
experiments have confirmed the 
violation of the Ellis-Jaffe Sum rule, and SMC and SLAC experiments have 
confirmed the Bj\"{o}rken Sum rule to about 10\% accuracy\cite{smcf,e143f,e154}.
The largest uncertainty that remains in the analyses now comes from the
low $x$ unmeasured region $x < 0.003$, which none of these experiments 
can access.  
The origin of the proton spin is still an unsolved mystery. Gluons in
the protons and the partonic angular moment may hold the answers.
The SLAC and SMC collaborations have strongly advocated future polarized DIS
measurements in the unmeasured low $x$ region\cite{smcf,e154}, while in near
future HERMES at DESY, COMPASS at CERN, and RHICSPIN at BNL will 
investigate the nucleon spin differently\cite{jaf99}.

In HERA we have a collider facility with $\sim 27.5$ GeV electron or 
positron beam, and a proton beam with $\sim 820$ GeV (in 1999 it has also been
operated with $E_{\rm p}$ = 925 GeV). With such facility one can access 
$\sqrt{s} \sim 300$ GeV.
These characteristics of the
HERA accelerator together with the H1 and ZEUS detectors enable 
measurements of the proton structure in the $x-Q^{2}$ 
range: $ \sim5 \times 10^{-5} \le x \le 0.6$ and 
$1 \le Q^{2} \le 5 \times 10^{5}$ (figure \ref{fig:xq2}, Bottom).
The electron beam is transversely polarized due 
to the Sokolov Ternov Effect (STE)\cite{sok64} during acceleration. 
Polarizations 
of $\sim 55\%$ have routinely been achieved at HERA\cite{barb} and 
are being used by the HERMES Collaboration\cite{herm} in their physics program. 
The proton beam is unpolarized. 
If it is polarized, measurements of spin variables could be possible  
in the extended kinematic range of HERA. This would be a 2 orders of magnitude 
increase
in the $x-Q^{2}$ range. This is precisely the
region identified by the SMC and the E154 collaborations where further
measurements are needed\cite{smcf,e154}.  

In this paper I will argue, based on the possible physics program 
with polarized $e-p$ collisions at HERA and the profound impact it may have
on our understanding of pQCD, that it would be unfortunate not 
to polarize the proton beam in HERA. 
The measurements at HERA will include spin structure function 
$g_{1}$ at low $x$, the 
measurement of polarized gluon distribution using three or four 
independent methods, each with different sources of experimental systematic 
uncertainties giving ample opportunity to check the results against each other,
measurement of parton distributions inside the polarized photon, semi-inclusive 
measurements to giving us access to the valence quark distributions, 
measurement of the weak structure functions $g_{5}$, weak interaction physics
at high $Q^{2}$, and possible physics beyond the Standard Model (SM) 
including lepto-quarks, contact interactions, and SUSY searches. I will 
make the case that, with the diverse physics measurements possible with 
the accelerator {\em and} 
the two collider detectors that already exist and operational, it would 
be unfortunate not to pursue the physics program with polarized protons 
at DESY. 

Dedicated studies on the physics topics presented in this paper started 
in a working group of the 1995/96 DESY workshop on 
{\em Future Physics at HERA}\cite{hera96} 
and were carried through in detail mainly in the 1997 DESY workshop on 
{\em Physics with Polarized Protons at HERA}(hence forth called ``the Workshop''
in this paper). The details of all the 
studies can be found in its proceedings\cite{pph97}.
Possible $\vec{p}_{\rm HERA}-\vec{p}_{\rm fixed-target}$ scattering, an option within 
the HERA-$N$ program, is discussed in \cite{now99}.
 
\section{Physics with Polarized Protons at HERA}
\label{sec:polhera}
Although the extension of $x-Q^{2}$ range of the unpolarized structure function
measurements has been the 
principal motivation for building the HERA accelerator and the two 
collider detectors, exciting new and unexpected physics has resulted
from the unpolarized DIS measurements\cite{derook99,caldwell99}.
Guided by this hindsight, while the major motivation for the measurement 
of spin variables at HERA is the measurement of spin structure function 
$g_{1}$ at low $x$ and the measurement of polarized gluon distribution,
in order to explore the full potential of the polarized HERA Collider 
other topics were also pursued in the Workshop\cite{pph97}. In this section 
only the most important topics out of a long list are reviewed. The following 
set of machine parameters and running conditions were assumed in most of 
the studies:\\
1. {\bf Integrated luminosity} over three to four years: ${\cal L} = 200 - 500$ pb$^{-1}$\\
$\Rightarrow$ Consistent with the $\sim 170$ pb$^{-1}$/year expected 
after the HERA luminosity upgrade in 2001.\\
2. {\bf Proton and electron/positron beam polarizations}: $P_{\rm p/e} = 70\%$\\
$\Rightarrow$ For electrons this is the design goal of HERA. While 55\% has 
been achieved routinely, up to 65\% has been achieved for short periods. 
For protons it is the goal with which the machine physics effort is being 
made\cite{barb97}.\\
3. {\bf Beam polarimetry}: relative uncertainty $\left[\delta (P_{\rm p/e})/ P_{\rm p/e} \right] \le 5\%$\\
$\Rightarrow$ For electrons this has already been achieved, while for protons
this remains a goal.\\
4. A {\bf fast detector simulation} of a HERA Collider
detector\cite{fds} was made available which 
could be used for measurability studies of different physical processes.

\subsection{Polarized Structure Functions of the Nucleons}
\label{sec:g1}
The measurement method for the spin structure function $g_{1}$ is well 
defined. The ratio of the difference between the event rate of  
$\vec{e}-\vec{p}$ scattering when the longitudinal spin vectors of the 
$\vec{e}$ and $\vec{p}$ are  parallel and when they are anti-parallel 
to the sum of the event rates is the measured asymmetry $A_{\rm m}$. 
$A_{\rm m}$ is used to evaluate the spin structure function $g_{1}$\cite{smcf} 
using the 
knowledge of kinematic factors such as the depolarization factor $D(y)$, 
the proton and electron polarizations $P_{\rm p/e}$, the unpolarized structure
function ratio $R = \sigma_{\rm T}/\sigma_{\rm L}$, and the unpolarized
proton structure function $F_{2}$.
\begin{figure}
\hfil
\epsfxsize=8.7cm
\epsffile{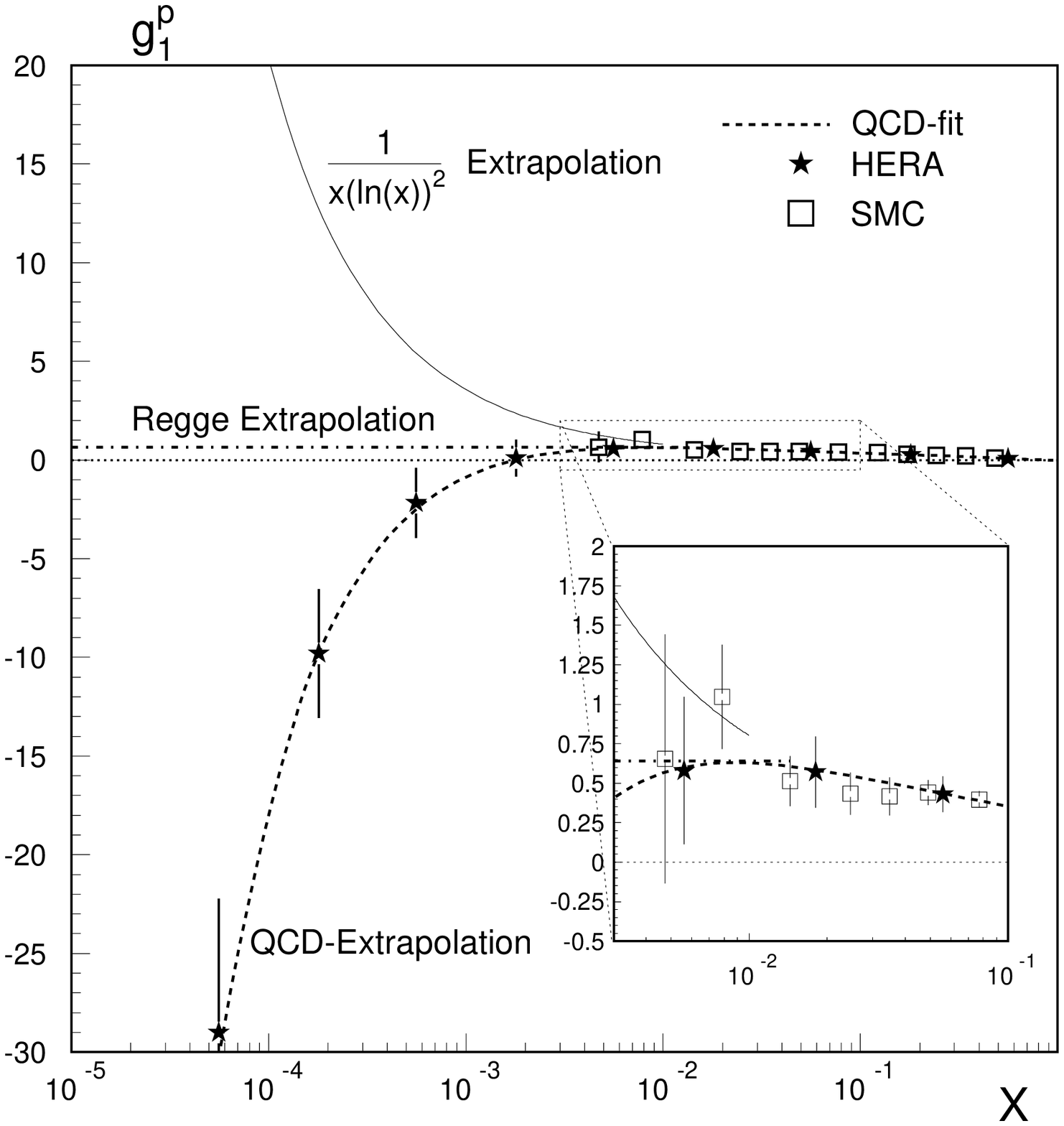}
\epsfysize=9.0cm
\epsffile{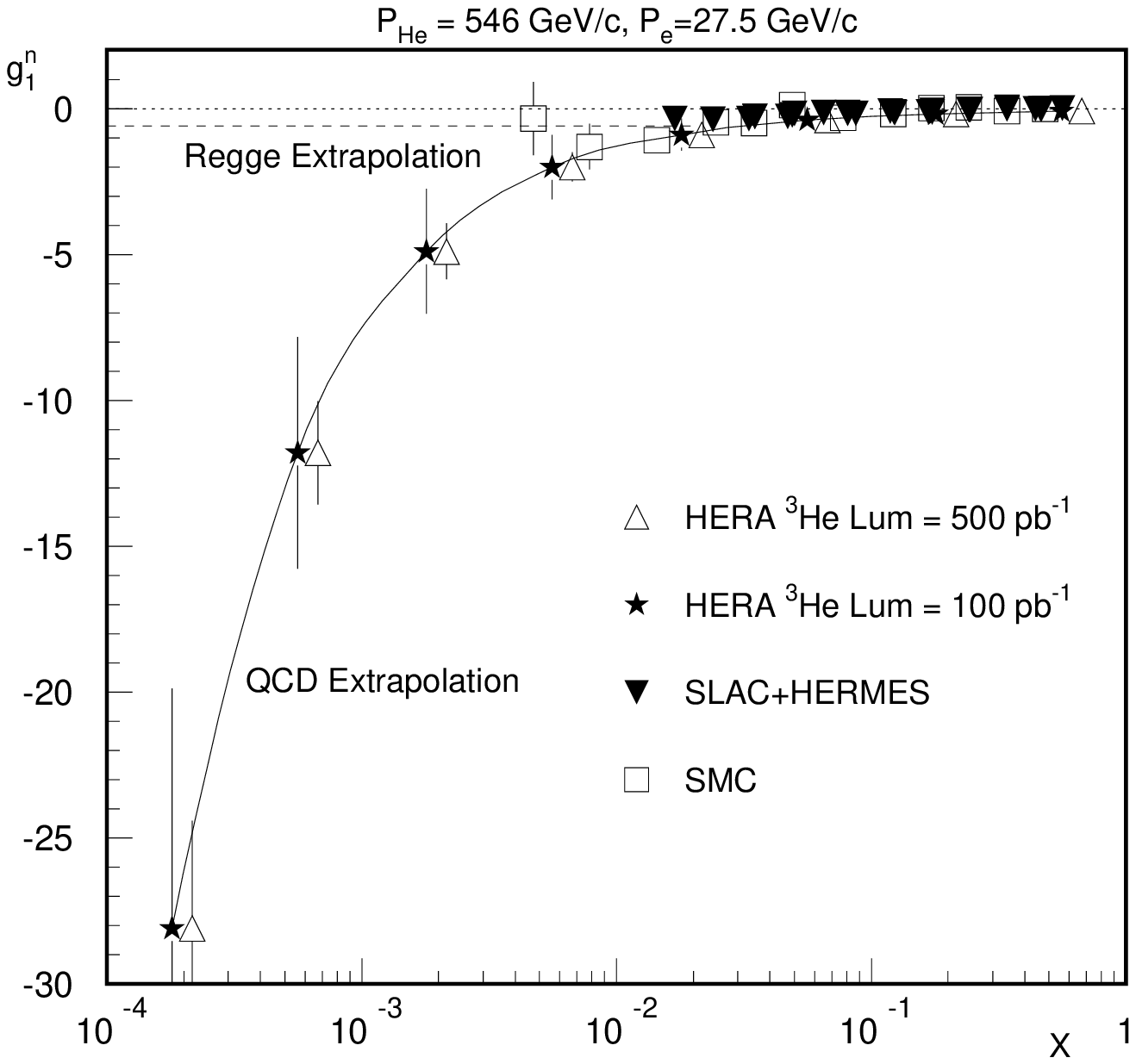}
\hfil
\caption{\em Top: The statistical uncertainty in the measurement of 
$g_{1}^{\rm p}$ possible at HERA shown with 
${\cal L}=500 {\rm pb}^{-1}$, 70\% polarization of beams
and using H1/ZEUS detectors. Bottom: The statistical accuracy of 
$g_{1}^{\rm n}$ measurement with $^{3}He^{+2}$ accelerated and stored 
in the proton ring of HERA. In both cases SMC and SLAC measurements 
are shown for comparison. All theoretical curves are drawn at $Q^{2}=10$ GeV.}
\label{fig:lowx}
\end{figure}
\subsubsection{Spin Structure of the Proton}
The possibility of \gp~measurement with the polarized HERA collider was studied
in detail\cite{ald97,epj99}. It was found to be one of the most important 
and unique measurement
that could be performed with the existing collider detectors at DESY.
Figure \ref{fig:lowx}(a) shows the statistical uncertainties in the measurement
of \gp~ possible with polarized HERA with ${\cal L} = 500~{\rm pb}^{-1}$, both
beam polarizations $70\%$, and using any one of the collider detectors at HERA.
Also shown in the figure are three different low-$x$ extrapolations, 
based on the available data from the fixed target experiments.
The HERA (pseudo)data are superimposed on the extrapolation of pQCD 
calculation based on a fit to fixed target data in the region 
$x > 0.003$ and $Q^{2} >1$ GeV$^{2}$.
The Regge inspired extrapolation 
$g_{1}(x\rightarrow 0) \sim x^{\alpha}; 0 \le \alpha \le 0.5$ is shown for
the value of $\alpha = 0$\cite{smcf,e143f}. A more exotic behavior with 
a functional form 
$1/\left[x({\rm ln}^{2}(x)) \right]$ proposed in \cite{close} is also shown. 
Not shown in this figure are low $x$ QCD predictions dominated
by double logarithmic resummations as suggested in \cite{kwi97}, which are similar
to the pQCD curve, but even more divergent. It is obvious from this 
figure that measurements with polarized HERA would easily 
resolve between these different theoretical scenarios. Since the underlying 
physics that drives these three behaviors is different, experimental 
confirmation of any of them would significantly enhance our understanding of the
low $x$ region. 

A subtle difficulty in this measurement is that the measured asymmetries 
predicted at low $x$ are very small $\sim {\rm few} \times 10^{-4}-10^{-3}$\cite{ald97,epj99}.
To explore the low $x$ region, one must keep the systematic
uncertainties smaller than the statistical errors. 
Two important sources have been looked into. 
First, the systematic uncertainties due to uncertainties in radiative 
corrections have 
been studied and were found to be small for the kinematic regions of our 
interest\cite{blum97}. Second, the effect of event migration due to the 
material present in the detectors and event reconstruction 
could be potentially disastrous if a large number of
events migrate and dilute or enhance the physics asymmetries. Using a
fast detector simulation\cite{fds} this was studied recently\cite{aid99}.
It was found that changes in the measured values of \gp~ were less than 
the $1\sigma$ statistical uncertainties indicated in Figure \ref{fig:lowx}(a). 
Spin structure function $g_{1}^{\rm p}$ could be reliably measured with
polarized HERA.
\subsubsection{Spin Structure of the Neutron}
If $^{3}He^{+2}$ can be accelerated in the HERA ring with sufficient
polarization and intensity, measurement of \gn~is possible\cite{ald97}. 
The measurement is further improved if the hadronic (proton) remanent 
the in the $\vec{e}-\vec{He}$ collision is tagged and thus the dilution factor
in the collision is improved. Figure \ref{fig:lowx}(b) shows results
of such measurement using 500 and 200 pb$^{-1}$ luminosity plotted
on the extrapolation based on the pQCD prediction in the region $x \le 0.003$
where no direct or indirect measurements exist for \gn.  Also shown is the 
Regge inspired extrapolations of the neutron structure function in this region.
This measurement would be especially important in view of the observations
by the E154 collaboration
of a possible divergent trend in \gn~at low $x$\cite{e154}.
Clearly, polarized HERA with helium ions in the proton ring would 
be of great experimental value. Further, in combination with the 
\gp~  measurements
at HERA, it would then be possible to improve upon the accuracy of experimental
verification of Bj\"{o}rken which presently is limited by the lack of
measurements in the $x < 0.003$ region\cite{smcf}.

\subsection{Polarized Gluon Distribution Function}
\label{sec:gluon}
In the framework pQCD the first moment of
the polarized gluon distribution, $\Delta G$, contributes to the proton spin. 
Presently all published results on polarized gluon distribution, 
$\Delta G(x,Q^{2})$, come from pQCD analyses of the $g_{1}^{\rm p,d,n}$ 
data\cite{smcf,e143f,e154}. 
In this section
I present the impact of polarized HERA on this method and also present 
other ways to determine $\Delta G(x,Q^{2})$ with polarized HERA.  
\subsubsection{$\boldmath{\Delta G(x,Q^{2})}$ from Perturbative QCD Analysis 
of $g_{1}$ and HERA}
Since gluon appears in the DGLAP equations at the Next-to-Leading-Order (NLO),
this method of determination of $\Delta G(x,Q^{2})$ is somewhat weak. However, it 
is the only method successfully employed to access the 
gluon distribution in the polarized DIS until now. 
The value of $\Delta G$ using
the world set of data at the time of the study
was $\Delta G \sim 1.0 \pm 0.3({\rm stat}) \pm 1.0({\rm theo})$\cite{ald97,epj99}.
The large theoretical uncertainty results from the fact that very little is 
known about the $g_{1}$\cite{smcf,epj99} and its evolution in the
low $x$ region, precisely where polarized HERA will make a difference.
To study the impact of polarized HERA on this measurement the
available data from fixed target and the possible future HERA data were
analyzed together\cite{ald97,epj99}. It was estimated that the statistical 
accuracy in the $\Delta G$ would be reduced by a factor $\sim 2$ and the 
theoretical uncertainty by a factor $\sim 3$. This is
a significant improvement in the determination of $\Delta G$ from pQCD 
analysis in NLO of the spin structure function data.
\subsubsection{Photon Gluon Fusion: Di-Jet and High-p$_{\rm T}$ Hadron Production}
The NLO analysis mentioned in the previous subsection allows the determination
of the first moment of the polarized gluon distribution, while the 
functional form of the polarized gluon distribution remains only losely 
determined. Further, it is biased
by the functional form of the initial parton distribution input into the 
analysis. The most promising way to get around this problem and to determine
the $\Delta G(x,Q^{2})$ is to make
measurements of physical processes in which the polarized gluon enters
into the interaction at Leading-Order (LO). One such process is the photon
gluon fusion process which gives rise to a $q\overline{q}$ pair in the
interaction. Depending on the energy of these quark pairs, either jets are
seen in the detector or the two jets hadronize and two oppositely charged 
hadrons with high transverse momenta ($p_{\rm T}$) are detected.  The rate 
asymmetry in the production of two jets (di-jet) or the two hadrons is 
related to the the polarized gluon distribution. 
\begin{figure}[here]
\epsfxsize=10.cm 
\hfil
\epsffile[15 270 535 550]{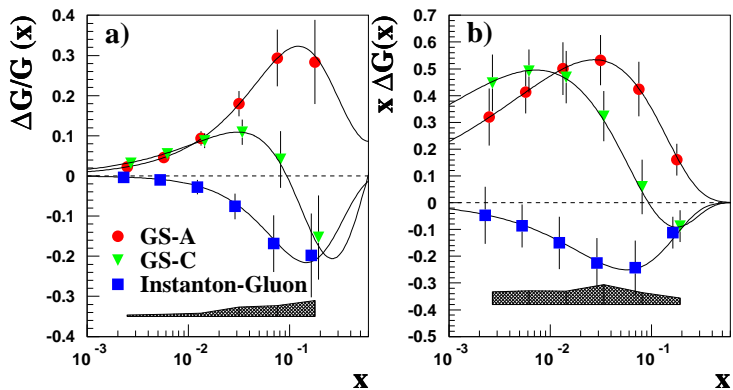}
\hfil
\caption{\em Statistical accuracy with polarized HERA in measurement of a) $\Delta G/G$ and b) x $\Delta G$ 
vs. $x$ using the asymmetry in di-jet production shown superposed on three different 
polarized gluon fit results.}
\label{fig:dijet}
\end{figure}
\begin{figure}
\epsfxsize=10.cm
\hfil
\epsffile[15 270 535 550]{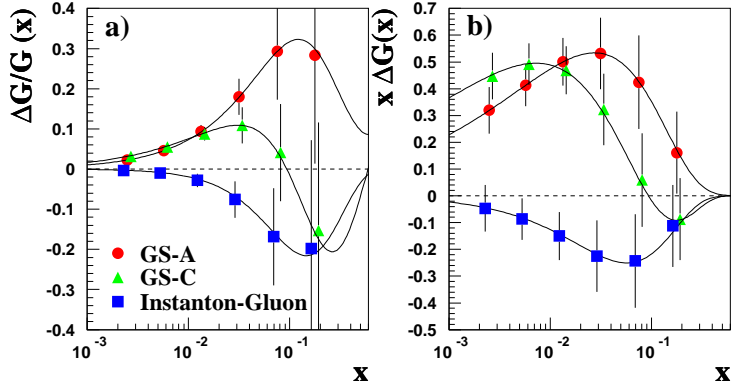}
\hfil
\caption{\em Same as in Fig.\ref{fig:dijet}, but determination based on 
two oppositely charged high-p$_{\rm T}$ hadron tracks.}
\label{fig:dihad}
\end{figure}

The study of what polarized HERA can contribute in this measurement was 
undertaken in the 1997 workshop\cite{gr97,gradr97}. Using LO theoretical
calculations, realistic detector acceptances and simulations it was shown
that polarized gluon distribution could be determined very accurately with
this method. Three different
but equally likely polarized gluon distributions, based on fits to
the available data {\em at the time}, are shown in Fig.
\ref{fig:dijet} for di-jets and Fig. \ref{fig:dihad} for di-hadrons. 
The possible statistical(error bars) are shown as well in the two figures.
The systematic(shaded horizontal bands) uncertainties are shown for
the di-jet (the band for di-hadron analysis is expected to be of the same
order). Clearly, the polarized HERA
will easily be able to distinguish between different scenarios. 

From the experimental point of view, an important advantage is that the 
two methods rely on different components of the the detector: di-jet on 
jet-detection and hadronic calorimeter, and charged hadron detection on 
tracking subdetectors. As such the systematic uncertainties are different.
This would then be independent methods to arrive at the polarized gluon
distribution. 

Although these calculations were done in LO at the time, recent advances
have enabled complete NLO calculations. The NLO calculations
do not affect the important the message of these studies, and further
re-inforce the fact that PGF would be a very reliable way to determine 
the polarized gluon distribution.
\subsubsection{Combined pQCD Analysis of $g_{1}$ and Di-Jet Events}
Perturbative QCD analysis of $g_{1}$ data
results in the first moment of the gluon distribution. It also gives
the shape of the distribution $\Delta G(x)$ at a fixed initial 
$Q^{2}_{0}$\cite{smcf,abfr}, but that functional 
form is input to the analysis and is only loosely determined. 
On the other hand the polarized gluon distribution determination via 
photon-gluon fusion process\cite{epj99} can constrain the shape of the distribution
better.
The limitation in this case is whether
complete range of $x_g$ is being explored by the accelerator/detector.
HERA, with its large beam energies and two large acceptance collider detectors,
provides the best opportunity to explore the largest possible $x_g$ 
range compared to any other experimental facility. The two methods combined 
(pQCD of $g_{1}$ and di-jet) clearly should completement each other and thus 
improve upon the resulting uncertainty in the determination 
$\Delta G(x,Q^{2})$ and its first moment.
It was shown in\cite{ald97,epj99} that the
power of such combined analysis is significant and a further reduction of
factor of 2 in the uncertainty of $\Delta G(x,Q^{2})$ and its first
moment is possible by combining the two analyses.


\subsection{Study of Photoproduction with Polarized HERA}
\label{sec:photop}
In the photoproduction limit, i.e. region where the intermediate
photon virtuality is small, the $e-p$ cross section can be approximated as
a product of photon flux factor and an interaction cross section of the
real photon with the proton. Measurements in this photoproduction
limit in the unpolarized physics program at HERA have not only 
led to significant improvement in our knowledge of the
structure of the proton and the photon, but also in the transition
from virtual to real photon. In the Workshop most of these physics topics
were explored assuming polarized proton and electron beams at HERA. 
The two most attractive results
are discussed in sections \ref{sec:photojets} and \ref{sec:dhgsum}.

Other topics such as open charm production, Drell-Yan processes,
large p$_{T}$ photon and inelastic $J/\Psi$ production were also studied. 
It was however concluded that some of them would need 
either higher luminosities than conservatively assumed in the Workshop 
or would need upgrades in the present detectors to achieve the required
efficiently for event tagging or detection. These upgrades although not
ruled out, would need further considerations in terms of detector upgrade.
As such, they are not included
in this paper, but the reader is referred to the proceedings \cite{pph97}.

\subsubsection{Jets and High-$p_{\rm T}$ Tracks in Photoproduction}
\label{sec:photojets}
\begin{figure}[t]
\hfil
\epsfysize=9.5cm
\epsffile{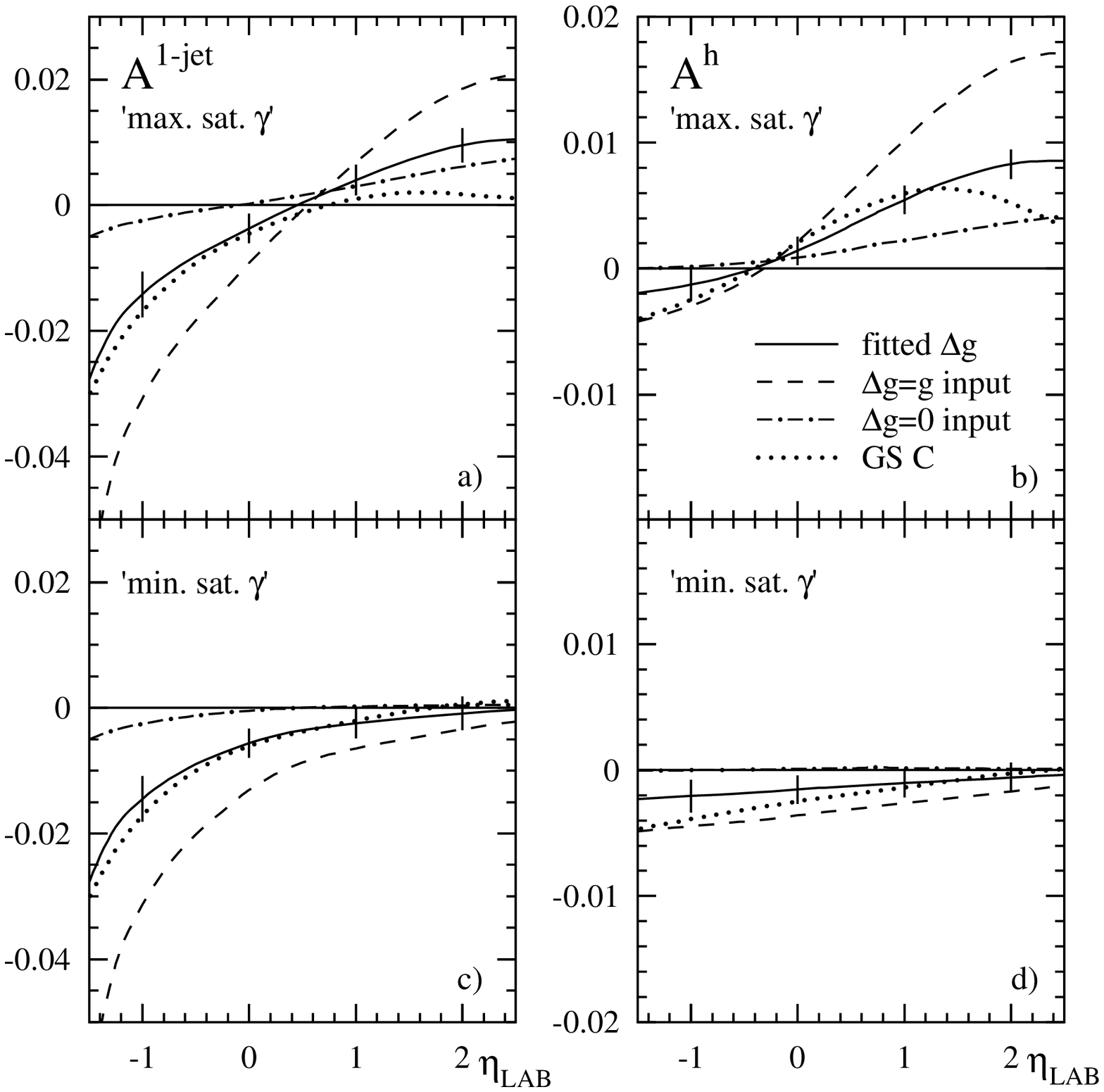}
\epsfysize=9.5cm
\epsffile{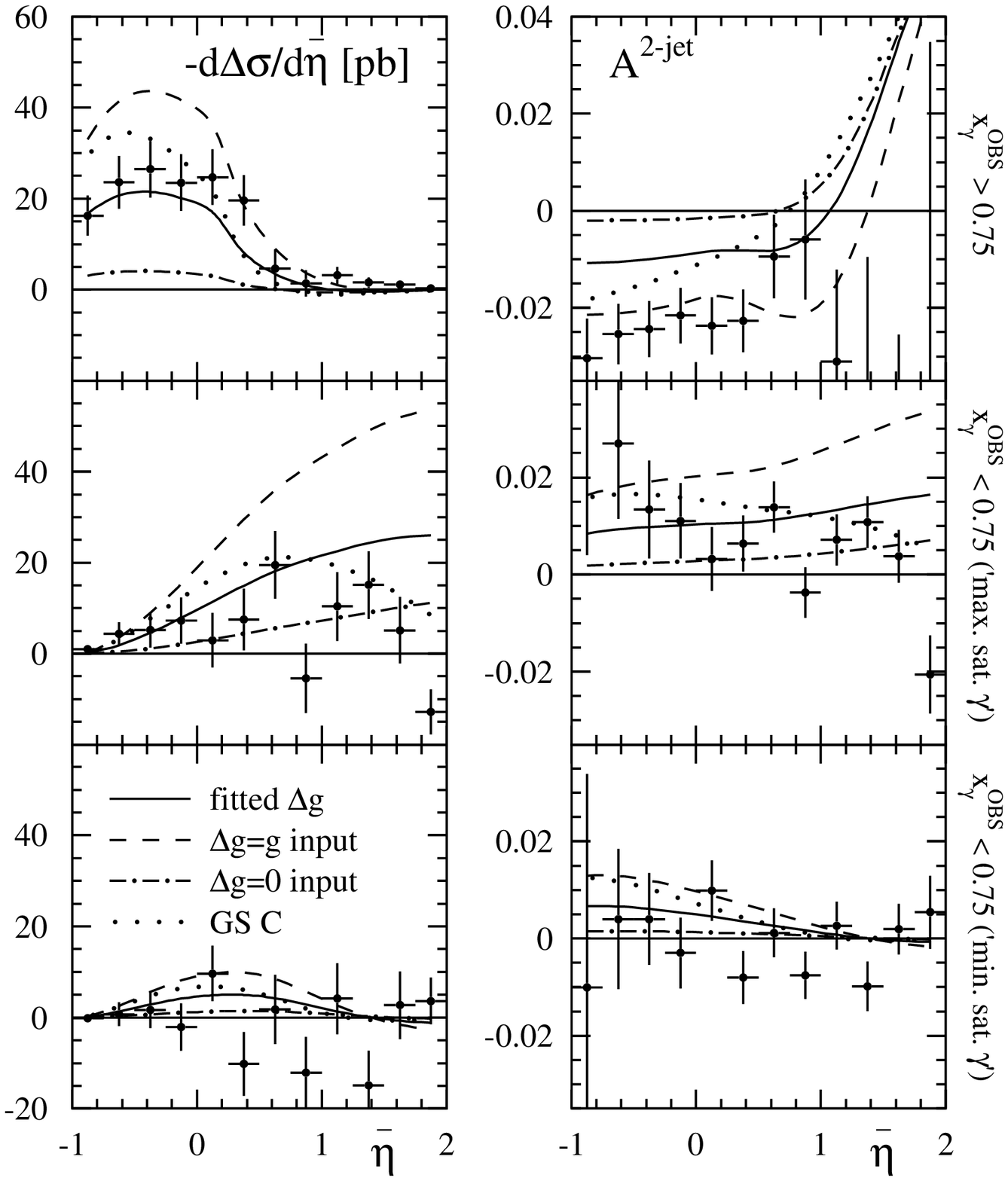}
\hfil
\caption{\em Statistical uncertainty in the measurement of asymmetry with
100 pb$^{-1}$ HERA luminosity for (a) single 
jet and (b) single hadron photoproduction while with 50 pb$^{-1}$ luminosity 
(c) differential cross section for
dijet photoproduction and (d) asymmetry in di-jet photoproduction. 
Different scenarios of polarized gluon distributions (indicated by the legends)
and possible scenarios regarding polarized photon distribution (top and bottom
in (a) and (b) and middle and bottom in (c) and (d)) are shown.
Max. Sat. $\gamma$ $\rightarrow \Delta \gamma = \gamma$ and Min. Sat. $\gamma$
$\rightarrow$ $\Delta \gamma = 0$.}
\label{fig:photop}
\begin{picture}(0,0)(0,10)
\put (60, 375){(c)}
\put (175,375){(d)}
\put (310,375){(a)}
\put (405,375){(b)}
\end{picture}
\end{figure}

A detailed study\cite{butt97} of physics with single jets and high-p$_{\rm T}$ 
tracks in polarized photoproduction including detector acceptance and standard
analysis cuts
was performed.
It showed that the study of  photoproduction of single inclusive jets as 
well as hadrons with polarized HERA would be an
extremely significant probe in investigating not only $\Delta G(x,Q^{2})$ 
but also the parton distribution inside the polarized photon, $\Delta q^{\gamma}$.
(See figures \ref{fig:photop}(a),(b)). With a modest luminosity
of 100 pb$^{-1}$, different theoretical scenarios for the
$\Delta G(x,Q^{2})$ as well as the $\Delta q^{\gamma}(x,Q^{2})$ could be resolved 
using the asymmetry in the jet and hadron production.
The single hadron production rate is somewhat lower than the jet production
rate, but it still could result in comparable measurements of $\Delta G$ 
and  $\Delta q^{\gamma}$ to those from jet production since less 
stringent constraints on the hadron track selections are warranted 
compared to the jet selection cuts. Although no quantitative estimates
of the improvement of uncertainty on $\Delta G(x,Q^{2})$ and 
$\Delta q^{\gamma}(x,Q^{2})$
were made, the figures clearly indicate that the statistical uncertainty 
being so small compared to the spread in different possible theoretical 
scenarios, the improvement in knowledge of polarized gluon and photon 
structure would indeed be significant.

A similar study for di-jet photoproduction was also taken up and resulted
in equally positive prospect of measurement of $\Delta G$ and $\Delta \gamma$.
Figure \ref{fig:photop} (c) and (d) show the statistical accuracy for di-jet
photoproduction with 50 pb$^{-1}$ HERA luminosity, standard analysis cuts,
and after a fast detector simulation, compared with different scenarios for the polarized gluon and
photon distribution. Clearly the different possible scenarios
are distinguishable from each other using the polarized HERA data in this case
as well. It also indicates that
the detector affects the measurement of the process minimally, as such, 
the measurement withstands the process of detection. Certainly these 
measurements performed with more luminosity will give significant information 
on the structure of the photon as well as the polarized gluon distribution.

While the measurement of structure of the polarized photon
is possible at other accelerator/detector facilities (LEP for example)
these measurements are not planned by the experimental collaborations.
The collisions in fixed target experiments are not energetic enough to 
access the photon structure. Measurements of the photon structure with 
polarized HERA will be unique and will provide valuable information.

\subsection{Inclusive Photoproduction Measurements}
\label{sec:dhgsum}
In section \ref{sec:photojets} we discussed the physics based on photoproduction
using only the detection of jets or high-p$_{T}$ hadrons.  Other topic which 
could be studied using photoproduction and can be characterized as 
``inclusive measurements'' is discussed below.

The H1 and ZEUS detectors routinely take data using the ``electron taggers''
situated in the beam pipe 6 to 45 meters downstream from the detectors. They
are used to measure the scattered electrons from events which have $Q^{2}$
in the range of $10^{-8} \rightarrow 10^{-2}$ GeV$^{2}$ and $\sqrt{s}$ in 
the range $ 60 \rightarrow 250$ GeV. The total cross section in this region
is $\sigma \sim 150 \mu$b. With these event characteristics the 
Drell-Hearn-Gerasimov (DHG) sum rule, which relates
the $\sigma^{\gamma p}_{\downarrow \uparrow}$ and  $\sigma^{\gamma p}_{\uparrow \uparrow}$,
to the anomalous magnetic moment $\kappa$ of the nucleon (proton in this case)
\begin{equation}
\int_{\nu_{\rm th}}^{\inf} \frac{{\rm d}\nu}{\nu} (\sigma^{\gamma p}_{\downarrow \uparrow} - \sigma^{\gamma p}_{\uparrow \uparrow})(\nu)= -\frac{4 \pi^{2} \alpha \kappa^{2}}{2m_{p}^{2}},
\end{equation}  
can be measured in the $\nu$ range of $10 - 40$ TeV\cite{bas97}. Although the 
contribution to the DHG integral from this region is small, a measurement at
HERA would be unique, because all other tests to check the
sum rule have been made using Regge inspired extrapolations of data 
from low energy fixed target experiments which explore $\nu$ values 
in the 10s of GeV range. Measurements at HERA will explore the energy 
dependence of the cross section in this yet unexplored region directly, 
and check the validity of the different Regge behaviors presently assumed.  

\subsection{Polarized Quark Distribution Functions}
\label{sec:g5}
Inclusive DIS measurements from $\gamma^{\star}$ exchange  
are sensitive to the sum of all quark flavors weighted by their charge 
squared. Using data from $e-p$ and $e-n$ scattering it is possible to extract singlet 
and non-singlet quark distributions, however, to extract separate quark
flavors additional information from semi-inclusive DIS measurements
is necessary. Ideally the measurement should be of the scattered hadron 
that carries the struck quark, but in practice the situation
is complicated because one can not avoid the effect of fragmentation
function and its convolution with various parton distributions. 

A detailed study of various issues relevant to semi-inclusive
DIS measurements at HERA was performed\cite{mau97}. Using the PEPSI Monte
Carlo to study the purity of flavored fragmentation functions for different
values of $z= P \cdot P_{h}/ P \cdot q$ revealed that at HERA purities
$\sim$90\% are possible at high values of $z$. As such, HERA is a good facility
to study the semi-inclusive scattering. Study of $\pi^{\pm}$ production
in low $x$ region showed that polarized valence and sea quark distributions 
could be measured with HERA luminosity $\ge 500$ pb$^{-1}$, but another
factor of 2 increase in the luminosity would be desirable.
\begin{figure}[here]
\epsfxsize=10.0cm
\hfil
\epsffile{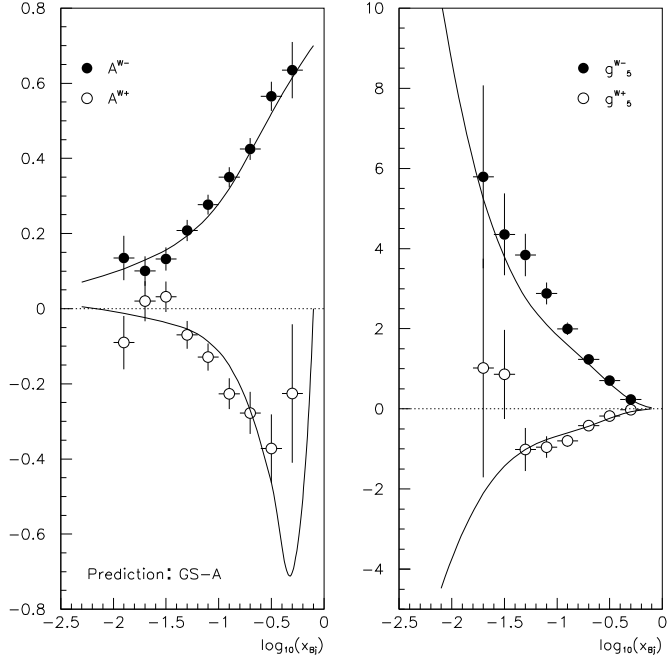}
\hfil
\caption{\em Left: Spin asymmetries $A^{W-}$ (full circles) and $A^{W+}$ (open 
circles) for cc-events and their statistical accuracy assuming 500 pb$^{-1}$ HERA
luminosity. Right: The structure functions $g_{5}^{W\pm}$ extracted from the
asymmetries. Detector effects move data points from the theoretical curves.}
\label{fig:g5}
\end{figure}

One can distinguish between the positively and negatively charged flavors
via $W^{\pm}$ exchange, i.e. via charged current (cc) interactions. It was
also shown in \cite{mau97} that with a modest luminosity of $\sim$ 
200 pb$^{-1}$, asymmetries are measurable for $W^{\pm}$ and using the pion
and kaon identification it would be possible to measure the relative contribution
to the spin from $\overline{s}$ and $\overline{d}$ quarks compared to the
$u$ quarks.

Using the data from charged current, one can also explore the partity violating
structure function $g_{5}^{W\pm}$. The asymmetry defined by
\begin{equation}
A^{W\pm} = \frac{d\sigma_{\uparrow \downarrow}^{W\pm} - d\sigma_{\uparrow \uparrow}^{W\pm}}
           {d\sigma_{\uparrow \downarrow}^{W\pm} + d\sigma_{\uparrow \uparrow}^{W\pm}}
         =\frac{\pm 2b g_{1}^{W\pm} + a g_{5}^{W\pm}}{aF_{1}^{W\pm} \pm bF_{3}^{W\pm}} \approx \frac{g_{5}^{W\pm}}{F_{1}^{W\pm}}
\end{equation}
where $a=2(y^{2}-2y+2)$ and $b=y(y-2)$ and $g_{5}^{W-}=\Delta u + \Delta c - \Delta \overline{d} -\Delta \overline{s}$ and
$g_{5}^{W+} = \Delta d + \Delta s - \Delta \overline{u} -\Delta \overline{c}$.
A Monte Carlo study including detector effects showed\cite{cont97} that the measurement
of the above asymmetry and hence the weak structure function $g_{5}$ was 
easily possible with the present detectors. The study concentrated on $Q^{2} \ge 225$ GeV$^{2}$.
Figure \ref{fig:g5} shows the result of the study. The statistical errors are
clearly small enough to make very good measurement of $A^{W\pm}$ and $g_{5}$.
The assumption made of course was that before the HERA is polarized a good
measurement of the unpolarized structure function $F_{1}^{W\pm}$ will
exist. From these measurements assuming that the contribution from $\Delta c$
is negligible, the distributions $\Delta u$ and $\Delta d$ could be extracted.

\subsection{High $Q^{2}$ Physics Outside The Standard Model}
\label{sec:hiq2}
Detection of neutral current events by the H1 and ZEUS collaborations at 
high $Q^{2}$ and high $x$ in excess compared to the Standard Model prediction
resulted in a great excitement two years ago\cite{hiq2}. These observations 
prompted considerable discussion in the particle physics community as possible
evidence for anomalies in the parton distributions of the proton or of physics
beyond the Standard Model. Since then both collaborations
have doubled their data sets but no new events have been found\cite{newhiq2}. 
As such, the statistical significance of the excess events originally observed is
now reduced. Although this is the case, there are other features that have
been observed in the data which, if confirmed with more statistical
significance, may 
shed light on physics within or outside of Standard Model. Most such 
effects are expected to have helicity specific properties.  What makes 
the investigations
at HERA even most interesting is the fact that some of these scenarios 
would result in situations in which neither Tevetron nor LEP2 data could 
resolve between different hypotheses, where as investigations with $e^{\pm}-p$ 
scattering could. Polarization of protons could add further information\cite{ell97} towards
our understanding of these outside-the-Standard Model scenarios.
Investigation of physics at high $Q^{2}$ with the polarized $e-p$ scattering 
will hence be important to understand the chiral structure of these 
interactions. Some specific cases investigated in the Workshop are 
discussed below.

Investigations of physics at high $Q^{2}$ with polarized protons and electrons
included the search for the evidence of R-Parity violating SUSY\cite{ell97},
Contact Interactions (CI)\cite{vir97}, and also search for the Instanton 
in protons\cite{koc97}. 
In a specific scenario: (leptoquarks) in R parity 
violating SUSY it was predicted\cite{ell97} that the \~{\em t} production 
cross section off of $d$ quark ($e\cdot d \rightarrow$ \~{\em t}) 
would result in a relation between cross sections: 
$\sigma(e^{+}_{R} p_{R}) \ll \sigma(e^{+}_{R} p_{L})$,
where $L,R$ indicate the left and right handedness of
the beams. Measurement of these cross sections would be crucial
to confirm these predictions and their underlying mechanism. 
If the \~{\em t} production happens 
through an $e\cdot s$ process, the cross sections were predicted to have
the following relations: $\sigma(e_{R}^{+}p_{R}) < \sigma(e_{R}^{+} p_{L})$ 
and $\sigma(e^{-}_{R} p_{L}) < \sigma(e^{+}_{L} p_{R})$. Again, polarized 
proton beams in HERA would be essential to measure these cross sections.

\begin{figure}[t]
\hfil
\epsfxsize=8.5cm
\epsffile[25 80 535 710]{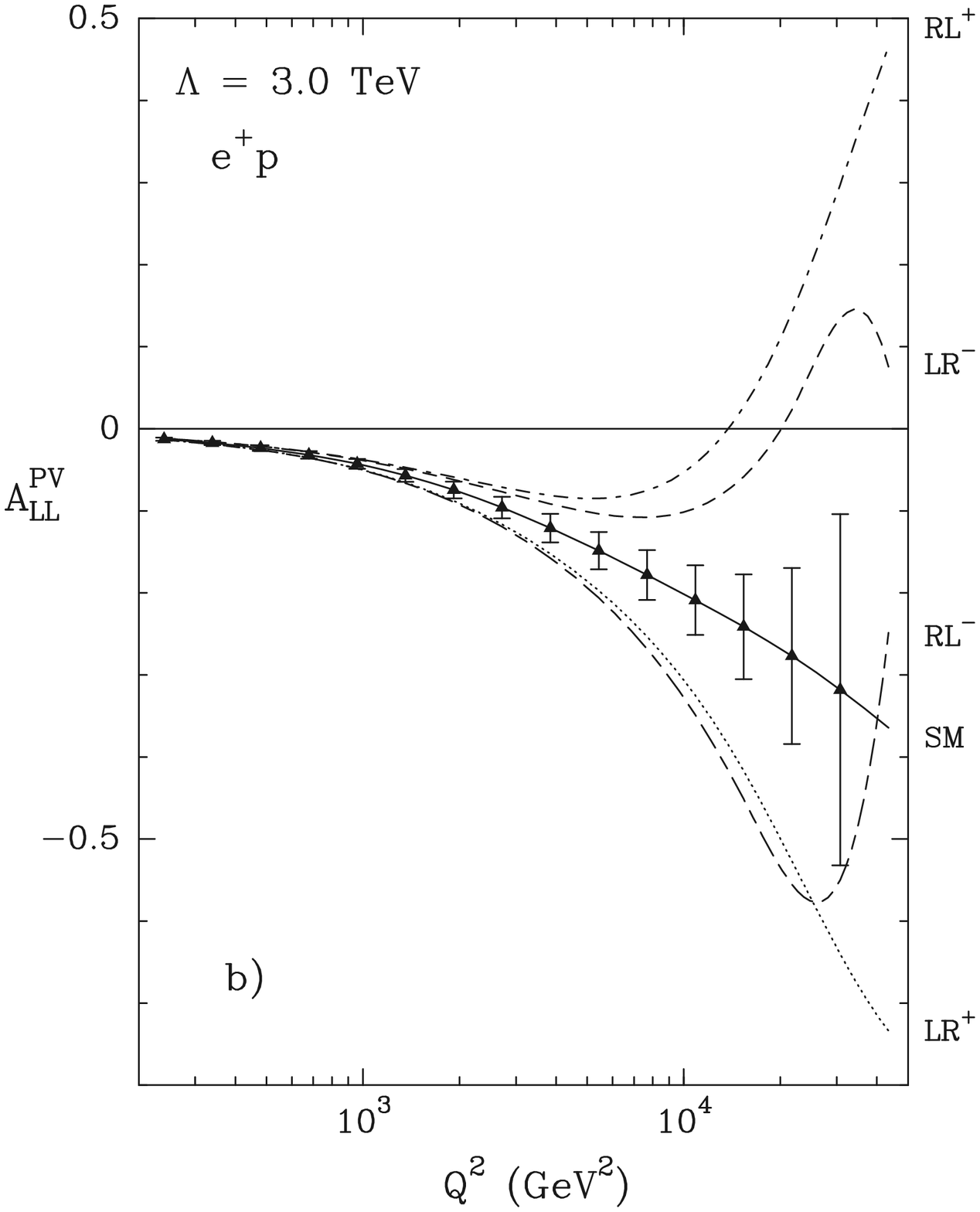}
\epsfxsize=8.5cm
\epsffile[25 80 535 710]{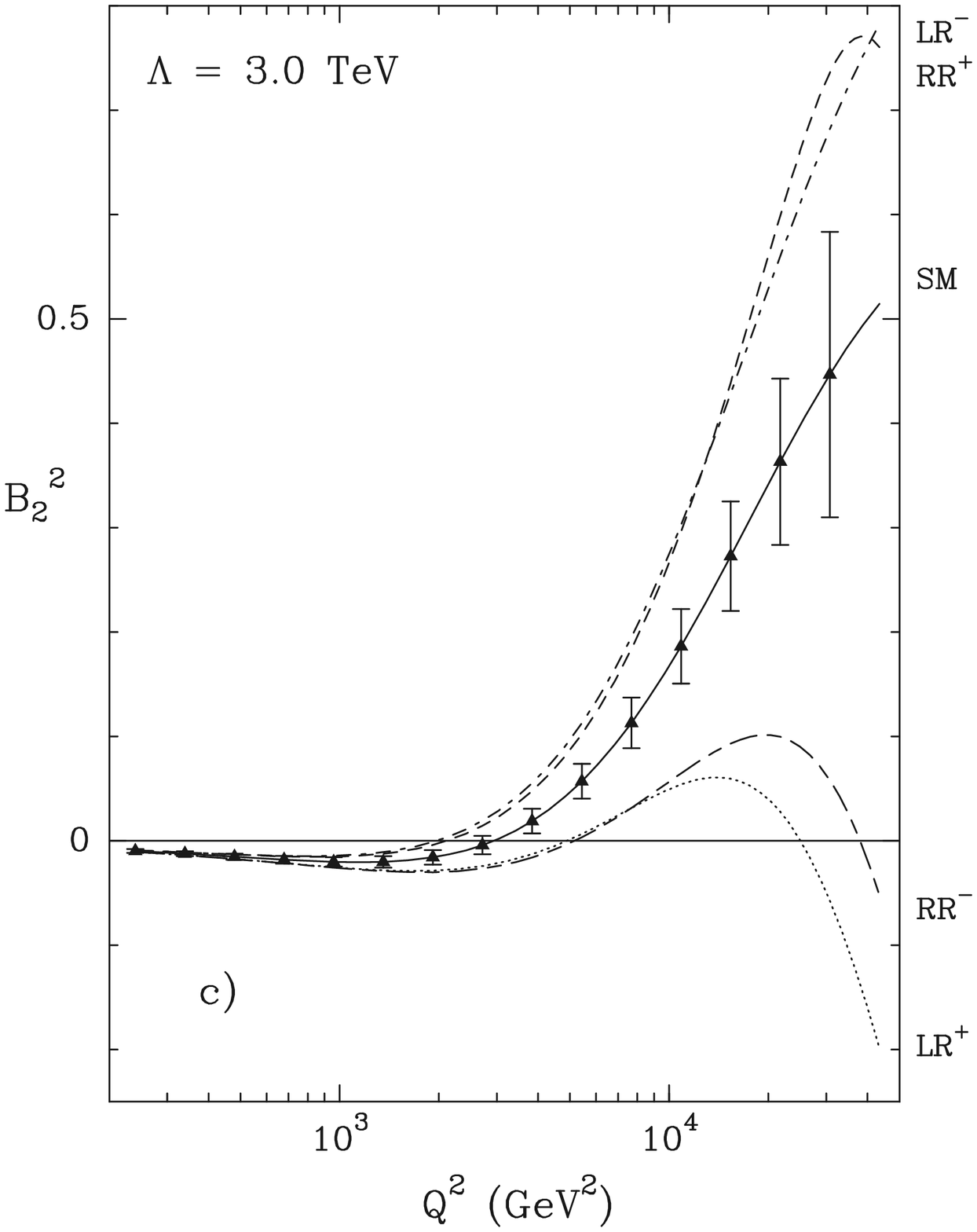}
\hfil
\caption{\em $A_{LL}^{\rm PV}$ and $B_{2}^{2}$ vs. $Q^{2}$. Different theoretical
scenarios shown with the possible statistical uncertainty with 250 pb$^{-1}$ 
luminosity for each of the $e^{\pm}$ and $\vec{e}-\vec{p}$ helicity combination. 
Standard Model prediction is shown for comparison.}
\label{fig:ci}
\end{figure}
The particle helicities and charges could play a most crucial role in the investigation
of possible Contact Interactions\cite{vir97} as well. Motivated by the present
results on the high $Q^{2}$ anomaly being related to possible Contact Interactions (CI),
predictions were made for specific asymmetries, some of them parity
violating, some parity conserving, while some of them mixed\cite{vir97}.
It was shown that if the CI effects should be visible with the HERA's 
kinematic reach at high $Q^{2}$, 
various types of CIs should manifest themselves in form of 
asymmetries {\em different} from those predicted by the Standard Model. 
The CI lagrangian is given by $$ {\cal L}_{CI} = \Sigma_{q=u,d} \eta_{i,j}(\overline{e}_{i} \gamma^{\mu} e_{i})
(\overline{e}_{j}\gamma_{\mu} e_{j})$$  where $\eta_{i,j}=(4\pi \epsilon)/\Lambda_{i,j},~\epsilon = \pm 1$ 
and $i,j$ stand for $L,R$, for left and right handedness. Including the
freedom to construct the asymmetries using these eight different types of
interactions and the differently charged $e$ beams and their helicities
along with the proton beam helicities, one can construct numerous
asymmetries. The most sensitive of these helicities were 
parity violating (PV) and mixed, are given by, 
\begin{equation}
A_{LL}^{\rm PV} = \frac{\sigma^{--}_{+} - \sigma^{++}_{+}}{\sigma^{--}_{+} + \sigma^{++}_{+}};~~~~
B_{2}^{2}(mixed) = \frac{\sigma^{++}_{-} - \sigma^{++}_{+}}{\sigma^{++}_{-} + \sigma^{++}_{+}}
\end{equation}
where the subscript indicates the charge of the electron beam, while the
superscript stands for the positive and negative helicities of colliding
beams. It was estimated that with the kinematics of the HERA accelerator
and the presently running detector acceptances and capabilities, the helicity
structure sensitivity up to $3-7$ TeV could be reached.
The asymmetries $A_{LL}^{PV}$ and $B_{2}^{2}$ estimated for different kinds
of contact interactions, as a function of $Q^{2}$ are shown in Figures \ref{fig:ci}(a) and
(b) respectively, and are compared with the Standard Model predictions 
along with the statistical accuracy that can be achieved with polarized HERA. 
The luminosity assumed in this study was 250 pb$^{-1}$ for each of 
the $e^{\pm}$ beam run with either helicity. Detector acceptances were 
included in this study. Apparently with polarized HERA the potential to 
indicate unambiguously if any deviation from the Standard Model occurs and 
to resolve some of it's helicity structure is good.

It was pointed out\cite{koc97} 
that in the Instanton Liquid Model a large effects from the 
instantons comes from high $Q^{2}$ and at high $x$. In stark contrast with
pQCD which predicts that the virtial photon-proton asymmetry 
$A_{1}^{\rm pQCD} \sim +1$ at high $x$, if instantons play a significant 
role in protons, the same asymmetry $A_{1}^{\rm Inst} \rightarrow -1$ as 
$x \rightarrow 1$ and at high $Q^{2}$. Measurements of $A_{1}$ at HERA 
should easily be able to resolve this.
\subsection{Spin and Fragmentation}
\label{sec:fragment}
The probability distribution of the proton fragmenting into a particular parton
in the polarized $e-p$ collisions is the polarized fragmentation function of
the proton. The measurement of polarized fragmentation functions is difficult
because this involves the measurement of the spin of the final state particles,
in fact, it is only possible in cases like $\Lambda$ baryon for which the
dominant decay mode is parity violating. Such studies, it should be noted,
do not need polarized protons beam. Only polarized electron beam can suffice.
As such these could start immediately after the electron spin rotators are
installed in the HERA ring during the luminosity upgrade. Different scenarios
for the spin transfer mechanism to $\Lambda$ baryon have been 
studied\cite{flo97_1,kot97, bur97}. It was demonstrated in \cite{flo97_1}
that the LEP data alone was {\em not} able to differentiate between
the different scenarios for the spin transfer assumed. However, with a moderate
luminosities of 100~pb$^{-1}$ $e-p$ scattering data at HERA, it would be
easy to disentangle these scenarios.

Introducing polarized protons in the above scheme enables possibilities
of new measurements. Fixed target polarized DIS experiments normally can 
access only the current fragmentation region. At HERA however one can 
easily study the current and the target fragmentation regions unambiguously
within the collider detectors. Polarized DIS at HERA will hence allow for 
the first time measurements of polarized target fragmentation. Of particular
interest is the scenario in which all the proton momentum would be 
detected in the target fragmentation region of the polarized DIS. In such
an eventuality the fracture function is expected to factorize into 
a transition probability of proton into a meson and another baryon, and the
structure function of the baryon. The experimental signature of this
scenario would be a highly energetic meson. Such measurements would lead us to
the structure functions of unstable baryonic excitations. If they are made
on different targets, i.e. with different hadron beams in the HERA ring,
one could measure the Ellis-Jaffe sum rule for different targets, and check if
the sum rule violation is target independent and indeed related to a
fundamental property of the QCD vacuum\cite{nar95}.

\section{Technical Challenges To Achieve Polarized HERA}
\label{sec:technical}
So far I have reviewed a few important topics out of the 
possible physics program that could be pursued with a polarized HERA collider. 
The breadth of the program and the fundamental nature of the problems it 
will address already make a compelling argument to do what it takes to achieve 
that goal. However, the argument in favor of polarized HERA, would not be 
complete without some comments the technical challenges facing the project, 
the details of these which are discussed elsewhere in these proceedings.
In this section I argue that although the goals from the
technical side {\em seem daunting}, the pace of advances in the recent past
has been impressive.
The final goal of a polarized HERA collider could be realized, after all,
if enough effort is made now.

\noindent{\bf High Luminosity:}
Presently, $\sim$ 35-40 pb$^{-1}$ luminosity is delivered by HERA
to the two collider detectors per year, and their efficiency to convert 
that into useful data is of the order of $\sim$ 70-80\%.
Most of the measurements discussed in section \ref{sec:polhera}
assumed typically 4 times higher luminosities than these.
A major effort is underway at DESY to increase the HERA luminosity by 
a factor of 4-5 to $\sim$ 170-200 pb$^{-1}$/year. This luminosity increase 
is essential for the electroweak and the high $Q^{2}$ physics program planned from 
2001-2005 at HERA\cite{cash96}. A polarized HERA program with such 
luminosity for $\sim$3-4years would be sufficient for most of the physics 
topics in section \ref{sec:polhera}.

\noindent{\bf Electron Beam Polarization and Polarimetry:}
The electron beam at HERA is naturally polarized.
Though not at its maximum achievable\cite{bar94} value $\sim$ 70\%  yet, 
$\sim$55\% has been routinely achieved\cite{herm} and occasional short 
periods with 65\% polarizations have been reported during dedicated 
accelerator tuning studies.
Effort towards achieving the maximum possible electron polarization 
are expected to continue through the luminosity upgrade program. 
Polarization measurement has also improved gradually. Starting
from $\delta P_{e}/P_{e} \sim$5.5\% in 1995\cite{herm} the best published 
value today is $\delta P_{e}/P_{e} \sim$ 3.4\% from 1997\cite{herm}. 
This is already at the level that is sufficient for the polarized HERA 
physics program in section \ref{sec:polhera}. 
In addition, there is a dedicated effort\cite{pol2000} to reduce the uncertainty 
further to $\le 2\%$ by year 2000.  

It is expected that the luminosity upgrade program at HERA will have some 
effect on the electron beam polarization due to beam-beam interactions in the
interaction region.  However, there is 
nothing to indicate that the $e$ beam polarization would be completely 
lost. It will have to be studied as the luminosity upgrade program 
proceeds.

\noindent{\bf Proton Beam Polarization and Polarimetry:}
Polarization of the proton beam is a challenging problem.
The protons being much heavier than electron, the synchrotron radiation and 
hence, the natural polarization due to STE\cite{sok64}, is almost negligible.
Realising a high energy polarized proton beam in a storage ring will require 
first, the development of a strong enough polarized H$^{-}$ source, followed by 
acceleration to high energy without destroying the beam polarization, 
and the finally, storage of the high energy polarized beam in the storage ring. 

The Relativistic Heavy Ion Collider (RHIC) at BNL plans to circulate 
polarized proton beams in the energy range 50-100 GeV by mid 2000 (eventually
250 GeV). They will be utilized for the RHIC-SPIN program\cite{rhicspin}. 
Although not quite at the very high energy of HERA, knowledge and 
experience gained in every aspect of the polarized proton beams at
RHIC would be invaluable towards achieving the polarized proton beam
at HERA.

A detailed study\cite{kri96} dedicated to polarized protons at HERA 
identified two major problems: a) adequately intense and highly polarized 
H$^{-}$ source and b) acceleration and retention of polarization in HERA.

HERA will need a polarized H$^{-}$ source of with the following 
specifications\cite{kri96,hof99}:
I=20 mA in 100 $\mu$s pulses at 0.25 Hz with emittance $2 \pi$ mm$\cdot$mr and
polarization $P \sim 80\%$. The most promising approach at the moment
seems to be the Optically Pumped Polarized Ion Source (OPPIS) development
at TRIUMF\cite{zel96}. Much progress has been made in the last
one year towards improving the source properties using this approach. 
The polarized H$^{-}$ source which will be used in the polarized proton
beam at RHIC is in its final stage of development at TRIUMF and is being brought
at BNL as this article goes to print. Studies towards achieving the HERA 
source specifications are under way\cite{kri96}.

Acceleration and storage of high energy polarized protons is difficult
mainly because of the depolarizing resonances. The large anomalous magnetic
moment of the proton ($\mu_{p}/\mu_{N} = 2.79$) results in high number of
spin precision per orbit (spin tune) equal to $G\gamma \sim 1530$ for 
HERA at 800 GeV\cite{hof99}. 
The spin motion
is thus very sensitive to the magnetic field, and in particular to the
imperfection and intrinsic resonances which can lead to depolarization
of the beam. Use of Siberian Snake Magnets (SSM) can avoid this depolarization. 
SSMs have been tested successfully recently at BNL for
the RHIC. 
Dedicated design study for HERA continues at DESY\cite{hof99}.

Proton beam polarimetry at high energy is being actively pursued at BNL.
Until recently no definite answers existed as to what could be used 
for the RHIC SPIN program. Developments in the last one year however
has changed this situation. Polarimeters
based on three different reactions are being considered\cite{bun99,hd99}: 
1) $\vec{p} \cdot C$ elastic scattering
in the Coulomb Nuclear Interference (CNI) region, 2) Inclusive pion production
$ \vec{p} \cdot C \rightarrow \pi^{+} X$, and 3) $\vec{p} \cdot p$ elastic
scattering using a {\em polarizable} jet target. The $\vec{p} \cdot C$ CNI 
and inclusive pion production polarimeters
were recently tested successfully at the AGS\cite{bun99}. Because of its low cost 
and possible high statistical precision in a short time, the CNI polarimeter 
became a natural choice for RHIC in its first year. It is being built now 
and expected to be ready by the end of 1999.

\section{A Case for Polarized HERA}
A future physics program at HERA with polarized proton and polarized $e^{\pm}$
beams would push study of the nucleon structure into an entirely new regime
using the {\em spin as a tool}, in addition to the high energy.
Not only are there clearly defined outstanding problems in the field 
which might be addressed by future data with polarized HERA, but it 
will also access possible {\em new} physics beyond the reach of any other 
accelerator facility, namely, the study of spin variables the low $x$ physics  
within the Standard Model, and the chiral structure of possible high 
$Q^{2}$ physics outside of it. History of this field has shown that  
exciting new results were obtained both in the unpolarized and the 
polarized DIS, every time a new $Q^{2}$ barrier was crossed. Polarized HERA will be
just such a step in polarized DIS. 

Most of the components
needed to pursue such a program already exist at DESY: $\vec{e}-p$ collider with sufficiently
high energy, polarized $e^{\pm}$ beam and a reliable measurement of the 
polarization, 
two working collider detectors H1 and ZEUS, and the people who have 
experience with both the machine and the detectors. The only missing 
component is the {\em polarized} proton beam.  

With so much already at hand it would be 
unfortunate not to make a determined pursuit of this goal with all 
the resources and efforts possible.

\noindent{\bf {Acknowledgement:}} It was a privilege to present this case on behalf of all the participants 
of the 1997 Workshop on Polarized Protons at HERA. Special thanks are due 
to A. De Roeck and G. R\"{a}del for their help in collecting all the material 
for the presentation. This work was supported by a grant from U.S. Department 
of Energy.

\end{document}